\documentclass[runningheads]{llncs}
\usepackage{graphicx}
\graphicspath{ {.} }
\usepackage{amsmath}
\usepackage{bookmark}
\usepackage{times}
\usepackage{xcolor}
\usepackage{verbatim}
\usepackage{xcolor}
\usepackage{soul}
\usepackage[utf8]{inputenc}
\usepackage[small]{caption}
\usepackage{hyperref}
\usepackage{listings}
\usepackage{enumitem}
\usepackage{amssymb}
\usepackage{csquotes}
\usepackage{subcaption}
\usepackage{algorithm}
\usepackage{algpseudocode}
\usepackage{newfloat}
\usepackage{float}

\DeclareFloatingEnvironment[placement={!ht},name=Listing, within=none]{instructionListing}
\DeclareMathOperator*{\argmax}{arg\,max}
\newcommand{\R}{\mathbb{R}}

\newcommand*{\sqspace}{\,}

\newcommand*{\psqqstring}{et sqq.}

\newcommand*{\psqq}{\sqspace\psqqstring}

\begin{document}
\title{Empathic Autonomous Agents}
%
%
\author{Timotheus Kampik\orcidID{0000-0002-6458-2252} \and
Juan Carlos Nieves\orcidID{0000-0003-4072-8795} \and
Helena Lindgren\orcidID{0000-0002-8430-4241}}
\authorrunning{T. Kampik et al.}
%
\institute{Umeå University, 901 87, Umeå, Sweden \\
\email{\{tkampik,jcnieves,helena\}@cs.umu.se}}
\maketitle              
\begin{abstract}
    Identifying and resolving conflicts of interests is a key challenge when designing autonomous agents.
    For example, such conflicts often occur when complex information systems interact persuasively with humans and are in the future likely to arise in non-human agent-to-agent interaction.
    We introduce a theoretical framework for an \textit{empathic} autonomous agent that proactively identifies potential conflicts of interests in interactions with other agents (and humans) by considering their utility functions and comparing them with its own preferences using a system of shared values to find a solution all agents consider \textit{acceptable}.
    To illustrate how empathic autonomous agents work, we provide running examples and a simple prototype implementation in a general-purpose programing language.
    To give a high-level overview of our work, we propose a reasoning-loop architecture for our empathic agent.

\keywords{Multi-agent systems \and Utility theory \and Conflicts of interests}
\end{abstract}
\section{Background and Problem Description}
In modern information technologies, conflicts of interests between users and information systems that operate with a high degree of autonomy (\textit{autonomous agents}) are of increasing prevalence.
For example, complex web applications persuade end-users, possibly against the interests of the persuaded individuals\footnote{E.g., research provides evidence that contextual advertisement influences how users process online news~\cite{doi:10.1080/0144929X.2016.1177115}; social network applications have effectively been employed for political persuasion (see for an example:~\cite{berinsky2017rumors}.}.
Given the prevalence of autonomous systems will increase, conflicts between autonomous agents and humans (or between different autonomous agent instances and types) can be expected to occur more frequently in the future, e.g. in interactions with or among autonomous vehicles in scenarios that cannot be completely solved by applying static traffic rules.
Consequently, one can argue for the need to develop \textit{empathic} intelligent agents that consider the preferences or utility functions of others, as well as ethics rules and social norms when interacting with their environment to avoid severe conflicts of interests.
As a simple example, take two vehicles (\textit{A} and \textit{B}) that are about to enter a bottleneck.
Assume they cannot enter the bottleneck at the same time.
A and B can either wait or drive.
Considering only its own utility function, A might determine that \textit{driving} is the best action to execute, given that B will likely stop and wait to avoid a crash.
However, A should ideally assess both its own and B's utility function and act accordingly.
If B's utility for driving is considered higher than A's, A can then come to the conclusion that \textit{waiting} is the best action.
As A does not only consider its own goals, but also the ones of B, one can regard A as \textit{empathic}, following Coplan's definition of empathy, as \textquote{a process through which an observer simulates another's situated psychological states, while maintaining clear self–other differentiation} \cite{doi:10.1111/j.2041-6962.2011.00056.x}.
While existing literature covers conflict resolution in multi-agent systems from a broad range of perspectives (see for a partial overview: \cite{10.1007/978-1-4020-6268-1_87}), devising a theoretical framework for autonomous agents that consider the utility functions (or preferences) of agents in their environment and use a combined utilitarian/rule-based approach to identify and resolve conflicts of interests can be considered a novel idea.
However, existing multi-agent systems research can be leveraged to implement core components of such a framework, as is discussed later.

In this chapter, we provide the following research contributions:
\begin{enumerate}
    \item We create a theoretical framework for an empathic agent that uses a combination of utility-based and rule-based concepts to compromise with other agents in its environment when deciding upon how to act.
    \item We provide a set of running examples that illustrate how the empathic agent works and show how the examples can be implemented in a general-purpose programing language.
    \item We propose a reasoning-loop architecture for a generic empathic agent.
\end{enumerate} 
The rest of this chapter is organized as follows: in Section~\ref{progress}, we present a theoretical framework for the problem in focus.
Then, we illustrate the concepts with the help of different running examples and describe the example implementation in a general-purpose programing language in Section~\ref{examples}.
Next, we outline a basic reasoning-loop architecture for the empathic agent in Section~\ref{architecture}.
In Section~\ref{jason}, we analyze how the architecture aligns with the belief-desire-intention approach and propose an implementation using the Jason multi-agent development framework.
Finally, we discuss how our empathic agent concepts relate to existing work, propose potential use cases, highlight a set of limitations, and outline future work in Section~\ref{discussion}, before we conclude the chapter in Section~\ref{conclusion}.

\section{Empathic Agent Core Concepts}
\label{progress}
In this section, we describe the core concepts of the empathic agent.
To allow for a precise description, we assume the following scenario\footnote{As we will explain later, the scenario and the resulting specification can be gradually extended to allow for better real-world applicability.}:
\begin{itemize}
    \item The scenario describes the interaction between a set of empathic agents $\{A_{0}, ..., A_{n}\}$.
    \item Each interaction scenario takes place at one specific point in time, at which all agents execute their actions simultaneously.
    \item At this point in time, each agent $A_i (0 \leq i \leq n)$ has a finite set of \textit{possible} actions ${Acts}_{i} := \{{Act}_{i}^0, ..., {Act}_{i}^m\}$, resulting in an overall set of action sets $Acts := \{{Acts}_{0}, ..., {Acts}_{n}\}$.
    Each agent can execute an \textit{action tuple} that contains one or multiple actions. In each interaction scenario, all agents execute their actions simultaneously and receive their utility as a numeric reward based on the actions that have been executed. 
    \item The utility of an agent $A_i$ is determined by a function $u_{i}$ of the actions of all agents.
    The utility function returns a numerical value or $null$\footnote{We allow for utility functions to return a $null$ value for action tuples that are considered impossible, e.g. in case some actions are mutually exclusive. While we concede that the elegance of this approach is up for debate, we opted for it because of its simplicity.}:
    \begin{align*}
        u_{i} := {Acts}_{0} \times ... \times {Acts_{n}} \rightarrow \{null, -\infty, \R, \infty\}
    \end{align*}
\end{itemize}
The goal of the empathic agent is to maximize its own utility as long as no conflicts with other agents arise.
We define a conflict of interests between several agents as any interaction scenario in which there is no tuple of possible actions that maximizes the utility functions of all agents.
I.e., we need to compare $\argmax u_{A_0}, ..., \argmax u_{A_n}$\footnote{The $\argmax$ operator takes the function it precedes and returns all argument tuples that maximize the function.}.
Note that $\argmax u_{A_i}$ returns a \textit{set of tuples} (that contains all action tuples that yield the maximal utility for agent $A_i$).
For this, we create a boolean function $c$ that the empathic agent uses to determine conflicts between itself and other agents, based on the utility functions of all agents:
\begin{align*}
    c&{}(u_{A_0}, ..., u_{A_n}) :=
    \left\{
        \begin{array}{ll}
            true, \: if: \\
             \quad \argmax u_{A_0} \cap ...  \cap \argmax u_{A_n} \ne \{ \}; \\
            false, \: otherwise.
        \end{array}
    \right.
\end{align*}
Considering the incomparability property of the von Neumann-Morgenstern utility theorem \cite{von1945theory}, such a conflict can be solved only if a system of values exists that is shared between the agents and used to determine \textit{comparable} individual utility values.
Hence, we introduce such a shared value system.
To provide a possible structure for this system, we deconstruct the utility functions into two parts:
\begin{itemize}
    \item An actions-to-consequences mapping (a function $a2c_i$ that takes the actions the agents potentially decide to execute and returns a set of \textit{consequences} (propositional atoms) $Consqs := \{{Consq}_{i}^0, ..., {Consq}_{i}^n\}$):
    \begin{align*}
        a2c_i := {Acts}_{i} \times ... \times {Acts_{n}} \rightarrow 2^{Consqs}
    \end{align*}
    \item A consequences-to-utility mapping (utility quantification function $uq$). Note that the actions-to-consequences mapping is agent-specific, while the utility quantification function is generically provided by the shared value system\footnote{I.e., for the same actions, an agent should only receive a different utility outcome than another agent if the impact on the two is distinguishable in its \textit{consequences}. We again allow for $null$ values to be returned in case of impossible action tuples.}:
    \begin{align*}
        uq := 2^{Consqs} \rightarrow \{null, -\infty, \R, \infty\}
    \end{align*}
\end{itemize}
Then, agents can agree on the utility value of a given tuple of actions, as long as the \textit{quality} of the consequence is observable to all agents in the same way.
In addition, the value system can introduce generally applicable rules, e.g. to hard-code a prioritization of individual freedom into an agent.
With help of the value system, we create a pragmatic definition of a conflict of interests as any situation, in which there is no tuple of actions that is regarded as \textit{acceptable} by all agents when considering the shared set of values, given each agent executes the actions that maximize their individual utility function.
To support the notion of \textit{acceptability}, we introduce a set of agent-specific \textit{acceptability functions} $accs := \{acc_{A_0}, ..., acc_{A_n}\}$.
The acceptability functions are derived from the corresponding utility functions and the shared system of values and take a set of actions as their inputs. Acceptability functions are domain-specific and there is no generic logic to be described in this context:
\begin{align*}
    acc_{A_i} := {Acts}_{A_0} \times ... \times {Acts}_{A_n} \rightarrow \{null, true, false\}
\end{align*}
The notion of acceptability rules adds a \emph{normative} aspect to the otherwise \emph{consequentialist} empathic agent framework.
Without this notion, our definition of a conflict of interests would cover many scenarios that most human societies regard as not conflict-worthy, e.g. when one agent would need to accept large utility losses to optimize its own actions towards improving another agents' utility.
Considering the acceptability functions, we can now determine whether a conflict of interests in terms of the \textit{pragmatic} definition approach exists for an agent $A_i$ by using the following function $cp$ that takes the utility function $u_i$ of agent $A_i$ and the acceptability functions $Accs := \{acc_{A_0}, ..., acc_{A_n}\}$ as input arguments:
\begin{align*}
    cp&{}(u_i, Accs) := \\
    &\left\{
        \begin{array}{ll}
            true, \: if: \\
            \quad \nexists acts \in \argmax u_{i} \land \forall_{acc \in Accs}: \: acc(acts) = true \\
            false, \: otherwise.
        \end{array}
    \right.
\end{align*}
We define an \textit{empathic} agent $A_i$ as an agent that, when determining the actions it executes, considers the utility functions of the agents it could potentially affect and maximizes its own utility only if doing so does not violate the acceptability function of any other agent; otherwise it acts to maximize the shared utility of all agents (while also considering the acceptability functions)\footnote{As different aggregation approaches are possible (for example: \emph{sum}, \emph{product}) to determine the maximal shared utility, we introduce the not further specified aggregation function $aggregate(u_{0}, ..., u_{n})$. In our running examples (see Section~\ref{examples}), we use the product of the individual utility function outcomes to introduce some notion of fairness; inequality should not be in the interest of the empathic agent. However, the design choice for this implementation detail can be discussed.}.
Algorithm~\ref{alg1} specifies an initial, \textit{naive} approach towards the empathic agent core algorithm.
The empathic agent core algorithm of an agent $A_i$ in its simplest form can be defined as a function that takes the utility functions $\{u_0, ..., u_n\}$ of the different agents, the set of all acceptability functions $Accs := \{acc_0, ..., acc_l\}$, and all possible actions $Acts_i$ of agent $A_i$ and returns the tuple of actions $A_i$ should execute\footnote{To facilitate readability, we switch to a pseudo-code notation for the following algorithms.}.
\begin{algorithm}[H]
    \caption{Naive empathic agent algorithm: $D\_A\_N$ (\emph{determine actions naive})}\label{alg1}
    \begin{algorithmic}[1]
        \Procedure{D\_A\_N$_i$}{$\{u_0, ..., u_n\}, Accs, Acts_i$} \Comment{Utility \& acceptability functions of all agents, actions of $A_i (0 \leq i \leq n)$}
            \If{$ \: \exists acts \in \argmax u_{i} \land \forall_{acc_{} \in Accs}: acc(acts) = true$}
            \State $best\_acceptable\_acts \gets \bigcup\limits_{acts \in \argmax u_{i}}: \forall_{acc \in Accs} acc(acts) = true$
            \State \textbf{return} $Acts_i \cap \: first(acts_k \in best\_acceptable\_acts)$
            \Else 
            \State \textbf{return} $Acts_i \cap \:  first(\argmax(aggregate(u'_{0}, ..., u'_{n}))$
            \EndIf 
        \EndProcedure
    \end{algorithmic}
\end{algorithm}
\noindent Note that in the context of the empathic agent algorithms, the function $first(set)$ turns the provided \emph{set of tuples} into a \emph{sequence of tuples} by sorting the elements in decreasing alphanumerical order and then returns the first element of the sequence.
This enables a deterministic action tuple selection.
Moreover, we construct a set of new utility functions $\{u'_0, ..., u'_n\}$ that assign all not acceptable action tuples a utility of $null$ (Algorithm~\ref{uprime})\footnote{We already use $null$ to denote \emph{impossible} action tuples. This implies an acceptable action tuple should always exists. To achieve a distinction, a value of $- \infty$ could be assigned.}:
\begin{algorithm}[H]
    \caption{Helper function: new utility function based on $u_i$; all not acceptable action tuples yield utility of $null$.}\label{uprime}
    \begin{algorithmic}[1]
        \Procedure{$u'_i$}{$u_i,\{{acts}_{0}, ..., {acts_{n}}\}, accs$}
        \State $is\_acceptable \gets \forall \: acc \in accs: acc({acts}_{i}) = true$
        \If{$is\_acceptable$}
        \State \textbf{return} $u_i({acts}_{i}, ..., {acts_{n}})$
        \Else
        \State \textbf{return} $null$
        \EndIf
        \EndProcedure
    \end{algorithmic}
\end{algorithm}
In Algorithm~\ref{alg1}, we specify that the agent picks the first item in the sequence of determined action tuples if it finds multiple optimal tuples of actions.
Alternatively, the agent could employ one of the following approaches to select between the optimal action tuples:
\begin{itemize} 
    \item \textbf{Random.} The agent picks a random action tuple from the list of the tuples it determined as optimal. This would require empathic agents to use an additional protocol to agree on the action tuple that should be executed.
    \item \textbf{Utilitarian.} Among the action tuples that were determined as optimal, the agent picks the one that provides maximal combined utility for all agents and falls back to a random or first-in-sequence selection between action tuples if several of such tuples exist.
\end{itemize}
Still, the algorithm is somewhat naive, as agents that implement it will decide to execute suboptimal activities if the following conditions apply:
\begin{itemize}
    \item Multiple agents find that the actions that optimize their individual utility are inconsistent with the actions that are optimal for at least one of the other agents.
    \item Multiple agents find that executing these \textit{conflicting} actions is considered \textit{acceptable}.
    \item Executing these \textit{acceptable} actions generates a lower utility for both agents than optimizing the shared utility would.
\end{itemize}
Hence, we extend the algorithm so that the agent selects the tuple of actions that maximizes its own utility, but falls back to maximize shared utility if the utility-maximizing action tuple is either not acceptable, or would lead to a lower utility outcome than maximizing the shared utility, considering the other agent follows the same approach (Algorithm~\ref{alglazy}):
\begin{algorithm}[H]
    \caption{Lazy empathic agent algorithm: $D\_A\_L$ (\emph{determine actions lazy})}\label{alglazy}
    \begin{algorithmic}[1]
        \Procedure{D\_A\_L$_i$}{$\{u_0, ..., u_n\}, Accs$} \Comment{Utility \& acceptability functions of all agents, actions of all agents $\{A_0, ..., A_n\}$}
            \State $\{acts\_max_0, ..., acts\_max_n\} \gets DETERMINE\_ACT\_MAX(u_{i}, Accs)$
            \State $\{good\_acts\_max_0, ..., good\_acts\_max_n\} \gets \{$ \\
                $\quad \quad \quad DETERMINE\_GOOD\_ACTS\_MAX(u_{0}, Accs, acts\_max_0),$ \\
                $\quad \quad \quad ...,$ \\
                $\quad \quad \quad DETERMINE\_GOOD\_ACTS\_MAX(u_{n}, Accs, acts\_max_n),$ \\
            $\quad \: \: \}$
            \If{$good\_acts\_max_0 \cap ... \cap good\_acts\_max_n \neq \{\}$}
            \State \textbf{return} $Acts_i \cap \:  first(good\_acts\_max)$
            \Else 
            \State \textbf{return} $Acts_i \cap \:  first(\argmax(aggregate(u'_{0}, ..., u'_{n})))$
            \EndIf 
        \EndProcedure
    \end{algorithmic}
\end{algorithm}
\noindent Algorithm~\ref{alglazy} calls two helper functions.
Algorithm~\ref{helpermaxacts} determines \emph{acceptable} action tuples that maximize a provided utility function $u_i$:
\begin{algorithm}[H]
    \caption{Helper function: determine acceptable action tuples that maximize utility function $u_i$}\label{helpermaxacts}
    \begin{algorithmic}[1]
        \Procedure{determine\_act\_max}{$u_{i}, Accs$}
        \State \textbf{return} $\bigcup\limits_{acts \in \argmax u_{i}}: \forall_{acc \in Accs} acc(acts) = true$
        \EndProcedure
    \end{algorithmic}
\end{algorithm}
\noindent Algorithm~\ref{helpergoodacts} determines all action tuples that would maximize an agent's ($A_i$'s) utility if this agent could \emph{dictate} the actions of all other agents, given the action tuples provide a better utility for this agent than the action tuples that maximize all agents' combined utility, given all agents execute an action tuple that maximizes their own utility if they could dictate the other agents' actions.
Note that Algorithm~\ref{helpergoodacts} makes use of the previously introduced algorithm (Algorithm~\ref{alg1}):
\begin{algorithm}[H]
    \caption{Helper function: determines all maximizing action tuples that would still yield a good utility result for agent $A_i (0 \leq i \leq n)$, given all other agents also pick an action tuple that would maximize their own utility, if all other agents \textquote{played along}.}\label{helpergoodacts}
    \begin{algorithmic}[1]
        \Procedure{Determine\_Good\_Acts\_Max}{$u_{i}, Accs$}
        \State \textbf{return} $\bigcup\limits_{acts \in \argmax u_{i}}:
        \forall_{acc \in Accs}: acc(acts) = true \: \land$ \\
        \quad \quad \quad \quad \: $u_i(\bigcup\limits_{k=0}^{n} D\_A\_N_k(\{u_0, ..., u_n\}, Accs, acts))$ \\ 
        \quad \quad \quad \quad $\geq u_i(acts\_max)$
        \EndProcedure
    \end{algorithmic}
\end{algorithm}
\noindent However, this algorithm only considers two types of action tuples for execution: action tuples that provide the maximal individual utility for the agent and action tuples that provide the maximal combined utility for all agents.
Action tuples that do not maximize the agent's individual utility, but are still preferable over the action tuples that maximize the combined utility, remain unconsidered. 
Consequently, we call an agent that implements such an algorithm a \textit{lazy} empathic agent.
We extend the algorithm to also consider all action tuples that could possibly be relevant.
I.e., if an action tuple is not considered acceptable, or if the tuple is considered acceptable but the agent chooses to not execute it, the agent falls back to the tuple of actions that provides the next best individual utility.
We construct a function $ne$ that returns the \textit{Nash equilibria} based on the updated utility functions $\{u'_0, ..., u'_n\}$, considering we have a strategic game $\langle N, (A_i) \succsim_i \rangle$, with $N := \{A_0, ... A_n\}$, $A_i := Acts_{A_i}$, and $acts \succsim_i acts' := u'_i(acts) \geq u'_i(acts')$\footnote{See the Nash equilibrium definition provided by Osborne and Rubinstein~\cite[p.~11\psqq]{osborne1994course}.}.
Then, we create the \textit{full} empathic agent core algorithm $D\_A\_F_i$ for an agent $A_i$ that takes the updated utility functions $\{u'_0, ..., u'_n\}$ and all agents' possible actions as inputs $\{{Acts}_{0}, ..., {Acts}_{n}\}$. The algorithm determines the (first of) the Nash equilibria that provide the highest shared utility and, if no Nash equilibrium exists, chooses the first tuple of actions that maximizes shared utility:
\begin{algorithm}[H]
    \caption{Full empathic agent algorithm: $D\_A\_F$ (\emph{determine actions full})}\label{full}
    \begin{algorithmic}[1]
        \Procedure{D\_A\_F$_i$}{$\{u'_0, ..., u'_n\}, \{{Acts}_{0}, ..., {Acts}_{n}\}$}
        \State $equilibria \gets ne(\{u'_0, ..., u'_n\}, \{{Acts}_{0}, ..., {Acts}_{n}\})$
        \If{$equilibria \neq \{\}$}
        \State $shared\_max\_equilibria \gets acts^* \in equilibria:$ \\
        \quad \quad \quad \quad $\forall acts \in equilibria:$ \\
        \quad \quad \quad \quad \quad $(u'_0(acts^*) \times ... \times u'_n(acts^*)) \geq (u'_0(acts) \times ... \times u'_n(acts))$
        \State \textbf{return} $Acts_i \cap \: first(shared\_max\_equilibria)$
        \Else \\
        \quad \quad \quad \textbf{return} $Acts_i \cap \: first(\argmax (aggregate(u_{0}, ..., u_{n}))$
        \State
        \EndIf
        \EndProcedure
    \end{algorithmic}
\end{algorithm}
Going back to the selection between several action tuples that might be determined as optimal, it is now clear that a \textit{deterministic} approach for selecting a final action tuple is preferable for both \textit{lazy} and \textit{full} empathic agents, as it avoids agents deciding upon executing action tuples that are not aligned with one another and lead to an unnecessary low utility outcome. Hence, we propose using a \textit{utilitarian} approach with a first-in-sequence selection if the utilitarian approach is inconclusive\footnote{As state above, we assume that the $first$ function sorts the action tuples in a deterministic order before returning the first element.}.

The proposed agent can be considered a \textit{rational agent} following the definition by Russel and Norvig in that it \textquote{acts so as to achieve the best outcome or, when there is uncertainty, the best expected outcome}~\cite[p.~4-5]{russell2016artificial} and an \textit{artificially socially intelligent agent} as defined by Dautenhahn as it instantiates \textquote{human-style social intelligence} in that it \textquote{manage[s] the individual's [its own] interests in relationship to the interests of the social system of the next higher level}~\cite{dautenhahn1998art}.

\section{Running Examples}
\label{examples}
In this section, we present two simple running examples of empathic agents and describe the implementation of the examples in a general-purpose programming language (JavaScript).

\subsection{Example 1: Vehicles}
\label{example1}
We provide a running example for the \textquote{vehicle/bottleneck} scenario introduced above.
Consequently, we have a two-agent scenario $\{A, B\}$.
Each agent has a utility function $u_{A,B}:= {Acts}_{A} \times {Acts}_{B} \rightarrow \{-\infty, \R, \infty\}$.
${Acts}_{A}$ and ${Acts}_{B}$ are the possible actions A and B, respectively, can execute.
To fully specify the utility functions, we follow the approach outlined above and first construct the actions-to-consequences mappings $a2c_A$ and $a2c_B$ for both agents.
The possible actions are $Acts_A = \{drive_A, wait_A\}$ and $Acts_B = \{drive_B, wait_B\}$. I.e., $Acts = \{drive_A, wait_A, drive_B, wait_B\}$.
To assess the consequences that include $waiting$, we assume $B$ is twice as fast as $A$ (without waiting, $A$ needs 20 time units to pass the bottleneck while $B$ needs 10)\footnote{$drive_A \land wait_A$ and $drive_B \land wait_B$, respectively, are mutually exclusive ($\{drive_A \oplus wait_A, drive_B \oplus wait_B\}$, with $A \oplus B := (A \lor B) \land \neg (A \land B)$). I.e., the functions return $null$ if $drive_A \land wait_A \lor drive_B \land wait_B$.}:
\begin{align*}
    a2c_A(acts) :=
    \left\{
        \begin{array}{ll}
            crash, \quad if: acts = (drive_A, drive_B);\\
            wait \: 0, \: \: \: if: acts = (drive_A, wait_B);\\
            wait \: \infty, \: if: acts = (wait_A, wait_B);\\
            wait \: 10, \: if: acts = (wait_A, drive_B);\\
            null, \: otherwise.
        \end{array}
    \right. \\
    \\
    a2c_B(acts) := \left\{
        \begin{array}{ll}
            crash, \quad if: acts = (drive_A, drive_B);\\
            wait \: 20, \: if: acts = (drive_A, wait_B);\\
            wait \: \infty, \: if: acts = (wait_A, wait_B);\\
            wait \: 0, \: \: \: if: acts = (wait_A, drive_B);\\
            null, \: otherwise.
        \end{array}
    \right.
\end{align*}
We construct the following utility quantification functions and subtract an amount proportional to the waiting time from the utility value $1$ of $wait \: 0$:
\begin{align*}
    u2c_A(consqs) := \left\{
        \begin{array}{ll}
            -\infty, \: if: consqs = \{crash\};\\
            0.9, \: \: \: if: consqs = \{wait \: 20\}; \\
            0 \quad \quad if: consqs = \{wait \: \infty\};\\
            1, \quad \quad if: consqs = \{wait \: 0\},\\
            null, \: otherwise.
        \end{array}
    \right. \\
    \\
    u2c_B(consqs) := \left\{
        \begin{array}{ll}
            -\infty, \: if: consqs = \{crash\};\\
            0.8, \: \: \: if: consqs = \{wait \: 20\}; \\
            0, \: \: \quad if: consqs = \{wait \: \infty\};\\
            1, \: \: \quad if: consqs = \{wait \: 0\};\\
            null, \: otherwise.
        \end{array}
    \right.
\end{align*}
Actions-to-consequences mappings and utility quantification functions can then be combined to utility functions:
\begin{align*}
    u_{A}(acts) ={}&
    \left\{
      \begin{array}{ll}
        \quad 1, \quad if: acts = (drive_A, wait_B); \\
    \: 0.9, \quad if: acts = (wait_A,  drive_B); \\
    \quad \ 0, \:\:\: if: acts = (wait_A, wait_B); \\
     -\infty, \:\: if: acts = (drive_A, drive_B); \\
    null, \:\:\: otherwise.  \end{array}
    \right. \\
    u_{B}(acts) ={}&
    \left\{
      \begin{array}{ll}
        \: 0.8, \quad if: acts = (drive_A, wait_B); \\
        \quad 1, \quad if: acts = (wait_A,  drive_B); \\
    \quad \ 0, \:\:\: if: acts = (wait_A, wait_B); \\
    \ -\infty, \:\: if: acts = (drive_A, drive_B); \\ 
    null, \:\:\: otherwise.
\end{array}
    \right.
\end{align*}
We assume scenarios where both agents are \textit{driving} or both agents are \textit{waiting} are not acceptable by either agents and introduce the corresponding acceptability rules:
\begin{align*}
    acc_{A,B}&{}(acts) :=
    \\&\left\{
        \begin{array}{ll}
            false, \: if: acts = (drive_A, drive_B) \lor (wait_A \land wait_B);\\
            null, \: if: (drive_A \in acts \land wait_A \in acts) \lor (drive_B \in acts \land wait_B \in acts);\\
            true, \: otherwise.
        \end{array}
    \right.
\end{align*}
Based on the utility functions ($u_A, u_B$), we create new utility functions ($u'_A, u'_B$) that consider the acceptability rules:
\begin{align*}
    u'_{A}(acts) :={}&
    \left\{
      \begin{array}{ll}
    \:\:\ \frac{1}{2}, \quad if: acts = (drive_A, wait_B); \\
    \:\:\ \frac{1}{3}, \quad if: acts = (wait_A, drive_B); \\
    null, \:\:\: otherwise.
  \end{array}
    \right. \\
    u'_{B}(acts) :={}&
    \left\{
      \begin{array}{ll}
        \:\:\ 1,  \quad if: acts = (wait_A, drive_B); \\
    \:\:\ \frac{1}{3}, \quad if: acts = (drive_A, wait_B); \\
    null, \:\:\: otherwise.  \end{array}
    \right.
\end{align*}
Finally, we apply the empathic agent algorithms to our scenario.
Using the \emph{naive} algorithm, the agents apply the acceptability rules, but do not consider the other agent's strategy. Hence, both agents decide to $drive$, (and consequently $crash$).
\begin{align*}
    D\_A\_N_A(\{u'_A, u'_B,\}, \{acc_A, acc_B\}, Acts_A) = drive_A \\
    D\_A\_N_B(\{u'_A, u'_B,\}, \{acc_A, acc_B\}, Acts_B) = drive_B
\end{align*}
The resulting utility is $-\infty$ for both agents.
None of the two other algorithms (\emph{lazy}, \emph{full}) allows any agent to decide to execute an action tuple that does not optimize shared utility.
I.e., both algorithms yield the same result:
\begin{align*}
    D\_A\_L_A(\{u'_A, u'_B,\}, \{acc_A, acc_B\}, Acts_A) = wait_A \\
    D\_A\_L_B(\{u'_A, u'_B,\}, \{acc_A, acc_B\}, Acts_B) = drive_B \\
\\
    D\_A\_F_A(\{u'_A, u'_B,\}, \{acc_A, acc_B\}, Acts_A) = wait_A \\
    D\_A\_F_B(\{u'_A, u'_B,\}, \{acc_A, acc_B\}, Acts_B) = drive_B
\end{align*}
The resulting utility is $0.9$ for agent A and $1$ for agent B.
As can be seen, the difference between agent types is not always relevant. The following scenario will provide a distinctive outcome for all three agent variants.

\subsection{Example 2: Concert}
\label{example2}
As a second example, we introduce the following scenario\footnote{The scenario is an adjusted and extended version of the \textquote{Bach or Stravinsky? (BoS)} example presented by Osborne and Rubinstein~\cite[p.~15--16]{osborne1994course}}.
Two empathic agents $\{A, B\}$ plan to attend a concert of music by either \emph{Bach}, \emph{Stravinsky}, or \emph{Mozart} ($Acts := \{Bach_A, Stravinsky_A, Mozart_A, Bach_B, Stravinsky_B, Mozart_B\}$).
A considers the Bach and Mozart concerts of much greater pleasure when attended in company of B (\emph{utility} of $6$, respectively $3$) and not alone (either concert: $1$).
In contrast, the Stravinsky concert yields good utility, even if A attends it alone ($4$). Attending it in company of B merely gives a utility bonus of $1$ (total: $5$).
B prefers concerts in company of A as well ($2$ for Stravinsky and $4$ for Mozart), but gains little additional utility from attending a Bach concert with A ($1.1$ with A versus $1$ alone) because they dislike listening to A's Bach appraisals.
Attending any concert alone yields a utility of $1$ for B.
As the utility is in this scenario largely derived from the subjective musical taste and social preferences of the agents and to keep the example concise, we skip the actions-to-consequences mapping and construct the utility functions right away\footnote{Note that the if-condition that triggers the return of a $null$ value simply defines that $Bach_A$, $Stravinsky_A$, and $Mozart_A$ are mutually exclusive, as are $Bach_B$, $Stravinsky_B$, and $Mozart_B$.}:
\begin{align*}
    u_{A}(acts) ={}&
    \left\{
      \begin{array}{ll}
        null, \:\:\: if: length(acts) \neq 2 \: \lor\\
        \quad \quad \quad length(set(Bach_A, Stravinsky_A, Mozart_A) \cap set(acts)) \neq 1 \lor\\
        \quad \quad \quad length(set(Bach_B, Stravinsky_B, Mozart_B) \cap set(acts)) \neq 1;\\
        \quad 6, \quad else \: if: acts = (Bach_A, Bach_B); \\
        \quad 5, \quad else \: if: acts = (Stravinsky_A,  Stravinsky_B); \\
        \quad 4, \quad else \: if: Stravinsky_A \in acts \land Stravinsky_B \notin acts; \\
        \quad 3, \quad else \: if: acts = (Mozart_A, Mozart_B); \\
        \quad 1, \quad otherwise.  \end{array}
    \right. \\
    u_{B}(acts) ={}&
    \left\{
      \begin{array}{ll}
        null, \:\:\: if: length(acts) \neq 2 \: \lor\\
        \quad \quad \quad length(set(Bach_A, Stravinsky_A, Mozart_A) \cap set(acts)) \neq 1 \lor\\
        \quad \quad \quad length(set(Bach_B, Stravinsky_B, Mozart_B) \cap set(acts)) \neq 1;\\
        \quad 1.1, \: else \: if: acts = (Bach_A, Bach_B); \\
        \quad 2, \quad else \: if: acts = (Stravinsky_A,  Stravinsky_B); \\
        \quad 4, \quad else \: if: acts = (Mozart_A, Mozart_B); \\
        \quad 1, \quad otherwise.
    \end{array}
\right.
\end{align*}
We introduce the following acceptability function that applies to both agents (although it is of primary importance for agent A).
As agent A is banned from the venue that hosts the Stravinsky concert, the action $Stravinsky_A$ is not acceptable:
\begin{align*}
    acc_{A,B}&{}(acts) :=
    \\&\left\{
        \begin{array}{ll}
            false, \: if: acts = Stravinsky_A \in acts;\\
            true, \: otherwise.
        \end{array}
    \right.
\end{align*}
Considering the acceptability function, we create the following updated utility functions:
\begin{align*}
    u'_{A}(acts) ={}&
    \left\{
      \begin{array}{ll}
        null, \:\:\: if: length(acts) \neq 2 \: \lor\\
        \quad \quad \quad length(set(Bach_A, Stravinsky_A, Mozart_A) \cap set(acts)) \neq 1 \lor\\
        \quad \quad \quad length(set(Bach_B, Stravinsky_B, Mozart_B) \cap set(acts)) \neq 1 \lor\\
        \quad \quad \quad Stravinsky_B \in acts; \\
        \quad 6, \quad else \: if: acts = (Bach_A, Bach_B); \\
        \quad 4, \quad else \: if: Stravinsky_A \in acts; \\
        \quad 3, \quad else \: if: acts = (Mozart_A, Mozart_B); \\
        \quad 1, \quad \:\:\: otherwise.  \end{array}
    \right. \\
    u'_{B}(acts) ={}&
    \left\{
      \begin{array}{ll}
        null, \:\:\: if: length(acts) \neq 2 \: \lor\\
        \quad \quad \quad length(set(Bach_A, Stravinsky_A, Mozart_A) \cap set(acts)) \neq 1 \lor\\
        \quad \quad \quad length(set(Bach_B, Stravinsky_B, Mozart_B) \cap set(acts)) \neq 1 \lor\\
        \quad \quad \quad Stravinsky_B \in acts; \\
        \quad 1.1, \: else \: if: acts = (Bach_A, Bach_B); \\
        \quad 4, \quad else \: if: acts = (Mozart_A, Mozart_B); \\
        \quad 1, \quad \:\:\: otherwise.
    \end{array}
\right.
\end{align*}
Now, we can run the empathic agent algorithms.
The \emph{naive} algorithm returns $Bach$ for agent $A$ and $Mozart$ for agent $B$:
\begin{align*}
    D\_A\_N_A(\{u'_A, u'_B,\}, \{acc_A, acc_B\}, Acts_A) = Bach_A \\
    D\_A\_N_B(\{u'_A, u'_B,\}, \{acc_A, acc_B\}, Acts_B) = Mozart_B
\end{align*}
The resulting utility is $1$ for both agents.
The \emph{lazy} algorithm returns $Mozart$ for both agents:
\begin{align*}
    D\_A\_L_A(\{u'_A, u'_B,\}, \{acc_A, acc_B\}, Acts_A) = Mozart_A \\
    D\_A\_L_B(\{u'_A, u'_B,\}, \{acc_A, acc_B\}, Acts_B) = Mozart_B
\end{align*}
The resulting utility is $3$ for agent A and $4$ for agent B.
The \emph{full} algorithm returns $Bach$ for both agents:
\begin{align*}
    D\_A\_F_A(\{u'_A, u'_B,\}, \{acc_A, acc_B\}, Acts_A) = Bach_A \\
    D\_A\_F_B(\{u'_A, u'_B,\}, \{acc_A, acc_B\}, Acts_B) = Bach_B
\end{align*}
The resulting utility is $6$ for agent A and $1.1$ for agent B.
\subsection{JavaScript Implementation}
\label{implementation}
\begin{figure}
    \includegraphics[width=\textwidth]{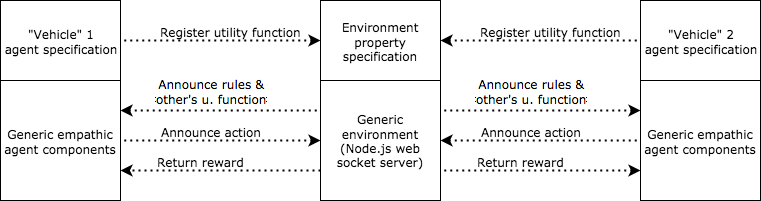}
    \caption{Empathic intelligent: architecture} \label{fig1}
\end{figure}

We implemented the running examples in JavaScript\footnote{The code, as well as documentation and tests, are available at \url{http://s.cs.umu.se/qxgbfi}.}.
As a basis for the implementation, we created a simple framework that consists of the following components:
\begin{itemize}
    \item \textbf{Web socket server: environment and communications manager.} The environment and communications interface is implemented by a web socket server that consists of the following components: 
        \begin{itemize}
            \item \textbf{Environment and communications manager.} The web server provides a generic environment and communications manager that relays messages between agents and provides the shared value system of acceptability rules.
            \item \textbf{Environment specification.} The environment specification contains scenario-specific information and enables the server to determine and propagate the utility rewards to the agents.
        \end{itemize}
    \item \textbf{Web socket clients: empathic agents.} The empathic agents are implemented as web socket clients that interact via the server described above.
        Each agent consists of the following two components:
        \begin{itemize}
            \item \textbf{Generic empathic agent library.} The generic empathic agent library provides a function to create an empathic agent object with the properties \textit{ID}, \textit{utilityMappings}, \textit{acceptabilityRules}, and \textit{type} (\textit{naive}, \textit{lazy}, or \textit{full}).
            The empathic agent object is then equipped with an action determination function that implements the empathic agent algorithm as described above.
            \item \textbf{Agent specifications.} The agent specification consists of the scenario-specific information of all agents in the environment, as well as of the current agents' identifier and type (\textit{naive}, \textit{lazy}, or \textit{full}) and is used to instantiate a specific empathic agent. Note that in the implementation, we construct the utility functions right away and do not use actions-to-consequences mappings.
        \end{itemize}
\end{itemize}
The implementation assumes that the specifications provided to both agents agents and to the server is consistent.
Fig.~\ref{fig1} depicts the architecture of the empathic agent JavaScript implementation for the \textit{vehicle} scenario.
We chose JavaScript as the language for implementing the scenario to show how to implement basic empathic agents using a popular general-purpose programing language, but concede that a more powerful implementation in the context of MAS frameworks like Jason is of value.

\section{Reasoning-loop Architecture}
\label{architecture}
\begin{figure}
    \includegraphics[width=\textwidth]{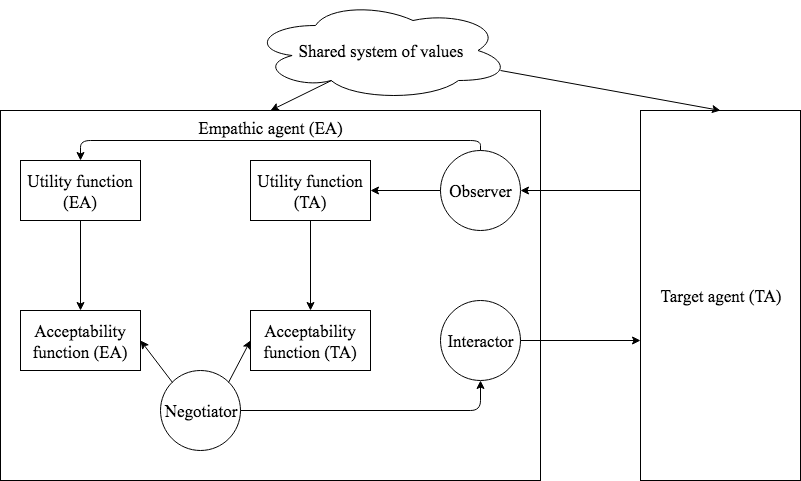}
    \caption{Empathic intelligent: architecture} \label{fig2}
\end{figure}

We create a reasoning-loop architecture for the empathic agent and again assume a two-agent scenario to simplify the description.
The architecture consists of the following components:
\begin{itemize}
    \item \textbf{Empathic agent (EA).}
    The empathic agent is the system's top-level component. It has three generic components (\textit{observer}, \textit{negotiator}, and \textit{interactor}) and five dynamically generated functions/objects (\textit{utility function} and \textit{acceptability function} of both agents, as well as a formalized model of the \textit{shared system of values}).
    \item \textbf{Target agent (TA).}
    In the simplest scenario, the empathic agent interacts with exactly one other agent (the \emph{target agent}), which is modeled as a black box.
    Pre-existing knowledge about the target agent can be part of the models the empathic agent has of the target agent's utility and acceptability functions.
    \item \textbf{Shared system of values.}
    The shared system of values allows comparing the utility functions of the agents and creating their acceptability functions, as well as their actions-to-consequences mappings and utility quantification functions, from which the utility functions are derived.
    \item \textbf{Utility function.}
    Based on the actions-to-consequences mappings and utility quantification functions, each empathic agent maintains its own utility function, as well as models of the utility function of the agent it is interacting with.
    \item \textbf{Acceptability function.}
    Based on the shared system of values, the agent derives the acceptability functions (as described above) to then incorporate them into updated utility functions, which it feeds into the empathic agent algorithm to determine the best possible tuple of actions.
    \item \textbf{Observer.}
    The observer component scans the environment, registers other agents, \emph{receives} their utility functions, and also keeps the agent's own functions updated.
    To construct and update the utility and acceptability functions without explicitly receiving them, the observer could make use of inverse reinforcement learning methods, as for example described by \cite{chajewska2001learning}. 
    \item \textbf{Negotiator.}
    The negotiator identifies and resolves conflicts of interests using the \textit{acceptability function} models and instructs the interactor to engage with other agents if necessary, in particular, to propose a solution for a conflict of interest, or to resolve the conflict immediately (depending on the level of confidence that the solution is indeed acceptable).
    The negotiator could make use of argument-based negotiation (see e.g.: \cite{Amgoud:2007:UGF:1329125.1329317}).
    \item \textbf{Interactor.}
    The interactor component interacts with the agent's environment and in particular with the target agent to work towards the conflict resolution.
    The means of communication is domain-specific and not covered by the generic architecture.
\end{itemize}
Fig.~\ref{fig2} presents a simple graphical model of the empathic agent's reasoning loop architecture.

\section{Alignment with BDI Architecture and Possible Implementation with Jason}
\label{jason}
Our architecture reflects the common belief-desire-intention (BDI) model as based on \cite{Bratman1987-BRAIPA} to some extent:
\begin{itemize}
    \item If a priori available to both agents in the forms of rules or norms, \textit{beliefs}, and \textit{belief sets} are part of the shared value system.
    Otherwise, they qualify the agents' utility and acceptability functions directly. In contrast, \textit{desires} define the objective(s) towards which an agent's utility function is optimized and are--while depending on beliefs--not directly mutable through persuasive argumentation between the agents.
    \item Intentions are the tuples of actions the agents choose to execute. 
    \item As it strives for simplicity, our architecture does for now not distinguish between desires and goals, and intentions and plans, respectively.
\end{itemize}
We expect to improve the alignment of our framework with the BDI architecture to facilitate the integration with existing BDI-based theories and implementation using BDI frameworks.
The Jason platform for multi-agent system development \cite{bordini2005bdi} can serve as the basis for implementing the empathic agent.
While simplified running examples of our architecture can be implemented with Jason, extending the platform to provide an empathic agent-specific abstraction layer would better support complex scenarios.

\section{Discussion}
\label{discussion}
In this section, we place our empathic agent concepts into the context of existing work, highlight potential applications, analyze limitations, and outline future work.

\subsection{Similar Conflict Resolution Approaches}
\label{comparison}
Our empathic agent can be considered a generic and basic agent model that can draw upon a large body of existing research on multi-agent learning and negotiation techniques for possible extensions.
A survey of research on agents that model other agents is provided by Albrecht and Stone \cite{albrecht_autonomous_2018}.
The idea of combining a utility-based approach with acceptability rules to emulate \textit{empathic} behavior is to our knowledge novel.
However, a somewhat similar concept is presented by Black and Atkinson, who propose an argumentation-based approach for an agent that can find agreement with one other agent on acceptable actions and can develop a model of the other agent's preferences over time~\cite{black2011choosing}.
While Black's and Atkinson's approach is similar in that it reflects Coplan's definition of empathy (it maintains \textquote{a process through which [it] simulates another's situated psychological states, while maintaining clear self–other differentiation}~\cite{doi:10.1111/j.2041-6962.2011.00056.x}) to some extent we identify the following key differences:
\begin{itemize}
    \item The approach is limited to a two-agent scenario.
    \item The agent model is preference-based and not utility-based. While this has the advantage that it does not require reducing complex preferences to a simple numeric value, it makes it harder to combine with existing learning concepts (see below).
    \item The agent has the ability to learn another agent's preferences over time. However, the learning concept is--according to Black and Atkinson--\textquote{not intended to be complete}~\cite{black2011choosing}. We suggest that while our empathic agent does not provide learning capabilities by default, it has the advantage that its utility-based concept allows for integration with established inverse reinforcement learning algorithms (see: Subsection~\ref{future}).
    \item The agent Black and Atkinson introduce is not \textit{empathic} in that it tries to compromise with the other agent, but rather uses its ability to model the agent's preferences to improve its persuasive capabilities by tailoring the arguments it provides to this agent.
\end{itemize}

\subsection{Potential Real-World Use Cases}
\label{use}
In this chapter, we exemplified the empathic agent with two simple scenarios, with the primary purpose of better explaining our agent's core concepts.
These scenarios do not fully reflect real-world use cases.
However, the core concepts of the agent can form the basis of solutions for real-world applications.
Below, we provide a non-exhaustive list of use case types empathic agents could potentially address:
\begin{itemize}
    \item \textbf{Handling aspects of traffic navigation scenarios that cannot be covered by static rules.} Besides adjusting the assertiveness levels to the preferences of their drivers, as suggested by Sikkenk and Terken \cite{Sikkenk:2015:RCA:2799250.2799270}, and Yusof et al. \cite{Yusof:2016:EAV:3003715.3005455}, autonomous vehicles could consider the driving style of other human- or agent-controlled vehicles to improve traffic flow, for example by adjusting speed or lane-changing behavior according to the (perceived) utility functions of all traffic participants or to resolve unexpected incidents (in particular emergencies).
    \item \textbf{Mitigating negative effects of large-scale web applications on their users.}
    Evidence exists that suggests the well-being of \textit{passive} (mainly content-consuming) users of social media is frequently negatively impacted by technology, while the well-being of at least some users, who actively engage with others through the technology, improves \cite{doi:10.1111/sipr.12033}.
    To facilitate social media use that is positive for the users' well-being, an empathic agent could serve as a mediator between user needs (social inclusion) and the business goals of the technology provider (often: maximization of advertisement revenue).
    \item \textbf{Decreasing the negotiation overhead for agent-based manufacturing systems.}
    Autonomous agent-based manufacturing systems are an emerging alternative to traditional, hierarchically managed control architectures \cite{MONOSTORI2006697}. While agent-based systems are considered to increase the agility of manufacturing processes, one disadvantage of agent-based manufacturing systems is the need for negotiation between agents and the resulting overhead (see for example: Bruccoleri et al. \cite{BRUCCOLERI2005433}). Employing empathic agents in agent-based manufacturing scenarios can possibly help solve conflicts of interests efficiently.
    \item \textbf{Improving persuasive healthcare technology.}
    Persuasive technology--\textquote{computerized software or information system designed to reinforce, change or shape attitudes or behaviours or both without using coercion or deception}~\cite{oinas2008towards}--is frequently applied in healthcare scenarios~\cite{conroy_behavior_2014}, in particular, to facilitate behavior change.
    Persuasive functionality is typically implemented using recommender systems~\cite{hors-fraile_analyzing_2018}, which in general struggle to compromise between system provider and end-user needs~\cite{ricci_recommender_2015}.
    This can be considered as a severe limitation in healthcare scenarios, where trade-offs between serving public health needs (optimizing for a low burden on the healthcare system) and empowering patients (allowing for a subjective assessment of health impact, as well as for unhealthy choices to support individual freedom) need to be made.
    Hence, employing the empathic agent concepts in this context can be considered a promising endeavor.
\end{itemize}

\subsection{Limitations}
\label{limitations}
The purpose of this chapter is to introduce \textit{empathic agents} as a general concept. When working towards a practically applicable empathic agent, the following limitations of our work need to be taken into account: 
\begin{itemize}
    \item The agent is designed to act in a fully observable world, which is an unrealistic assumption for real-world use cases.
    For better applicability, the agent needs to support probabilistic models of the environment, the other agents, and the shared value system.
    \item Our formal empathic agent description is logic-based. Integrating it with Markov decision process-based inverse reinforcement learning approaches is a non-trivial endeavor, although certainly possible.
    \item In the example scenarios we provided, all agents are identically implemented empathic agents.
    An empathic agent that interacts with non-empathic agents will need to take into account further game-theoretic considerations and to have negotiation capabilities.
    \item The presented empathic agent concepts use a simple numeric value to represent the utility an agent receives as a consequence of the execution of an action tuple.
    While this approach is commonly employed when designing utility-based autonomous agents, it is an oversimplification that can potentially limit the applicability of the agent.
    \item Software engineering and technological aspects of empathic agents need to be further investigated.
    In particular, the implementation of an empathic agent library using a higher-level framework for multi-agent system development, as we discuss in Section~\ref{jason} could provide a more powerful engineering framework for empathic agents.
\end{itemize}

\subsection{Future Work}
\label{future}
We suggest the following research to address the limitations presented in Subsection~\ref{limitations}:
\begin{itemize}
    \item So far, we have chosen a logic-based approach to the problem in focus to allow for a minimalistic problem description with low complexity.
    Alternatively, the problem could be approached from a \textit{reinforcement learning} perspective (see for an overview of multi-agent reinforcement learning: \cite{busoniu2008comprehensive}).
    Using (partially observable) Markov decision processes, one can introduce a well-established temporal and probabilistic perspective\footnote{However, the same can be achieved with temporal and probabilistic logic.}.
    A key capability our empathic agent needs to have is the ability to learn the utility function of other agents.
    A comprehensive body of research on enabling this ability by applying \textit{inverse reinforcement learning} exists (for example: \cite{chajewska2001learning} and \cite{ng2000algorithms}).
    Hence, creating a Markovian perspective on the empathic agent to enable the application of reinforcement learning methods for the observational learning of the utility functions of other agents can be considered relevant future work.
    \item To better assess the applicability of the empathic agent algorithms, it is important to analyze its computational complexity in general, as well as to evaluate it in the context of specific use cases that might allow for performance-improving adjustments.
    \item To enable empathic agents to reach consensus in case of inconsistent beliefs \\ argumentation-based negotiation approaches can be applied that consider uncertainty and subjectivity (e.g. \cite{Marey2015}) for creating solvers for finding compromises between utility/acceptability functions.
    Similar approaches can be used to enhance utility quantification capabilities by considering preferences and probabilistic beliefs.
    \item The design intention of the architectural framework we present in Section~\ref{architecture} is to form a high-level abstraction of an empathic agent that is to some extent agnostic of the concepts the different components implement. We are confident that the framework can be applied in combination with existing technologies to create a real-world applicable empathic agent framework, at least for use cases that allow making some assumptions regarding the interaction context and protocol.
    \item The ultimate goal of this research is to apply the concept in a real-world scenario and evaluate to what extent the application of empathic agents provides practically relevant benefits.
\end{itemize}

\section{Conclusion}
\label{conclusion}
In this chapter, we introduced the concept of an \textit{empathic agent} that proactively identifies potential conflicts of interests in interactions with other agents and uses a mixed utility-based/rule-based approach to find a mutually acceptable solution.
The theoretical framework can serve as a general purpose model, from which advanced implementations can be derived to develop socially intelligent systems that consider other agents' (and ultimately humans') welfare when interacting with their environment.
The example implementation, the reasoning-loop architecture we introduced for our empathic agent, and the discussion of how the agent can be implemented with a belief-desire-intention approach provide first insights into how a more generally capable empathic agent can be constructed.
As the most important future research steps to advance the empathic agent, we regard the conceptualization and implementation of an empathic agent with learning capabilities, as well as the development of a first simple empathic agent that solves a particular real-world problem. 
%
%
%
\subsubsection{Acknowledgements}
We thank the anonymous reviewers for their constructive critical feedback.
This work was partially supported by the Wallenberg AI, Autonomous Systems and Software Program (WASP) funded by the Knut and Alice Wallenberg Foundation.
\bibliographystyle{splncs04}
\bibliography{references}

\end{document}